\documentstyle[12pt]{article}

\def\eql{\stackrel{\rm law}{=}}
\newcommand{\gT}{\Theta}
\title{Models of Passive and Reactive Tracer Motion: an Application of 
Ito Calculus}
\author{Paolo Muratore Ginanneschi}
\newtheorem{theorem}{Theorem}
\begin{document}
\maketitle
\centerline{PDC KTH 100 44 Stockholm, Sweden \footnote{new address from 
$01/03/1997$: CATS-NBI, Blegdamsvej 17, DK-2100 Copenhagen, Denmark}}
\vfill
\begin{abstract}
By means of Ito calculus it is possible to find, in a straight-forward 
way, the analytical solution to some equations 
related to the passive tracer transport problem in a velocity field
that obeys the multidimensional Burgers equation and to a simple model 
of reactive tracer motion.
\end{abstract}
\vfill
\pagebreak

\section{Introduction}
\label{s:intro}

In a recent paper \cite{Saichev} Saichev and Woyczynski have obtained
exact solutions in arbitrary dimensions of equations of hydrodynamic 
type related to the Burgers equation (see, e.g., \cite{Burgers}, \cite{Frisch},
\cite{Saichev2}, \cite{VeDuFrNo}) for an irrotational velocity field with 
two models of coupled passive or reactive tracers.  For the inhomogeneous 
Burgers equation and more general models of tracers, closed but less explicit 
solutions are obtained as path integrals.

The main idea of \cite{Saichev} is to reduce the system specified by the 
forced Burgers equation together with the advection-diffusion-reaction 
equation, to a pair of coupled linear diffusion equations with variable 
coefficient which can be analytically solved by means of the Feynman-Kac 
formula (see, e.g.\cite{kara}).

They also show that the same methodology can be used to generate exact 
solutions of a non linear reaction diffusion equation coupled with a
Burgers-like velocity field also depending on the concentration.

The Feynman-Kac equation expresses the solution of a parabolic PDE 
without drift in terms of a conditional average over Brownian
trajectories. A generalization of the Feynman-Kac formula when a 
drift is present, is supplied by the Came\-ron-Mar\-tin-Gir\-sanov 
formula \cite{gich}. This observation allows to recover the 
results in \cite{Saichev} without using any auxiliary field.

\section{The Girsanov formula}
\label{s:girs}

In this section a generalized version of the Girsanov formula is recalled
and its relevance to parabolic PDEs is explained.

\begin{theorem}
Let $\vec x^{\,(1)}$, $\vec x^{\,(2)} \in {\bf R}^{d}$ be solutions on 
the interval $0\,\leq t\,\leq T$  of the stochastic differential equations
\begin{eqnarray}
d\vec x^{\,(i)}(t)&=&\vec b^{\,(i)}(\vec x^{\,(i)}(t),t)dt
+\sigma(\vec x^{\,(i)}(t),t)
d\vec w(t) \nonumber \\
\vec x^{\,(i)}(0)&=&\vec x\,\,\,\,\,\,\,\,\,\,i=1,2
\label{sdd}
\end{eqnarray}
where $\vec w(t)$ is a d-dimensional Brownian motion, 
$\vec b^{\,(i)}(\vec x,t)$ ($i=1,2$) and $\sigma(t,\vec x)$ are 
respectively Borel-measurable, ${\bf R}^{d}$-valued functions on 
$[0,\infty[\times{\bf R}^{d}$ and a Borel-measurable, 
$[0,\infty[\times{\bf R}^{d}\times{\bf R}^{d}$-valued function with bounded 
inverse $\forall \,\vec x\,,t$. 
If $\,\vec b^{\,(i)}(\vec x,t)$ ($i=1,2$) and $\sigma(t,\vec x)$ satisfy the 
assumptions of the existence and uniqueness theorem for the solutions of 
(\ref{sdd}), then the probability measure $\mu_{2}$ of $\vec x^{\,(2)}$ will 
be absolutely continuous w.r.t $\mu_{1}$ of $\vec x^{\,(1)}$ and
\begin{eqnarray}
\frac{d\mu_{2}}{d\mu_{1}}(\vec x^{\,(1)}(t))&=&e^{{\cal A}_{t}} \nonumber \\
{\cal A}_{t}&=&\int_{0}^{t}\vec \alpha(\vec x^{\,(1)}(s),s)\cdot 
d\vec w(s)-\frac{1}{2}\int_{0}^{t}\|\vec \alpha(\vec x^{\,(1)}(s),s)\|^{2}\,ds
\label{girsanov}
\end{eqnarray}
where
\begin{equation}
\vec \alpha(\vec x,t)=\sigma^{-1}(\vec x,t)\,[\vec b^{\,(2)}(\vec x,t)
-\vec b^{\,(1)}(\vec x,t)\} \nonumber \\
\end{equation}
\end{theorem}
see \cite{gich} pag. 279 for proof and details.

A straight-forward consequence is that for any reasonably smooth function
$\it{f_{k}(x)}$, with k ranging from $1$ to $n$  and for all n-tuples
$(t_{1},\cdots,t_{n})$ such that $0\,\leq\,t_{1}\,\cdots\,\leq\,t_{n}
\,\leq\,t$ we have:
\begin{equation}
E^{\vec x}\{\prod_{k=1}^{n}\it{f_{k}(\vec x^{\,(2)}(t_{k}))}\}=E^{\vec x}\{
\prod_{k=1}^{n}\it{f_{k}(\vec x^{\,(1)}(t_{k}))}e^{{\cal A}_{t}}\}
\end{equation}
 
This result can be exploited in order to write the solutions of parabolic 
PDEs as path integrals on Wiener trajectories (see e.g. \cite{kara}).

In the following sections the result (\ref{girsanov}) will be applied, 
disregarding the conditions on the drift field, in order to derive formally 
the announced results. The advantage of this approach is to supply a 
direct physical interpretation for the solutions in terms of stochastic 
trajectories.

\section{Simple applications}
\label{s:simple}

As a first application let as consider the d-dimensional homogeneous 
Burgers equation with rotation free initial condition:
\begin{eqnarray}
\partial_{t}\vec v +\vec v \cdot\vec \nabla\vec v&=&\nu\Delta\vec v
\nonumber \\
\vec v(\vec x,0)&=&\vec \nabla \gT_{0}(\vec x) 
\label{buro}
\end{eqnarray}
The physical meaning of (\ref{buro}) is that the velocity field is, 
on the average, constant along the trajectories generated by the stochastic 
differential equation:
\begin{eqnarray}
d\vec x(s)&=&-\vec \nabla \gT(\vec x(s),t-s)\,ds+\sqrt{2\,\nu}\,d\vec w(s) 
\nonumber \\
\vec x(0)&=&\vec x 
\label{eqdifstobu}
\end{eqnarray}
If we introduce:
\begin{eqnarray}
d\vec z(s)&=& \sqrt{2\,\nu}\,d\vec w(s) \nonumber\\
\vec z(0)&=&\vec x  \nonumber\\
\vec z(t)&\eql&{\cal N}(\vec x,2\,\nu\,t) 
\label{zeta}
\end{eqnarray}
we can exploit Girsanov's theorem and write:
\begin{equation}
\vec \nabla \gT(\vec x,t)=E^{\vec x}\{\vec \nabla_{z(t)}\gT_{0}(\vec z(t))
e^{-{\cal Z}_{t}}\}
\label{scasol}
\end{equation} 
where:
\begin{equation}
{\cal Z}_{t}=\frac{1}{\sqrt{2\,\nu}}\int_{0}^{t}\vec \nabla 
\gT(\vec z(s),t-s)\cdot d\vec w(s)+\frac{1}{4\,\nu}\int_{0}^{t}\|\vec \nabla 
\gT(\vec z(s),t-s)\|^{2}\,ds
\label{expsto}
\end{equation}
We can eliminate the stochastic integral in (\ref{expsto}) by means of
\begin{eqnarray}
\lefteqn{d_{s}\gT(\vec z(s),t-s)=}\nonumber\\
&=&\{-\partial_{t-s}\gT(\vec z(s),t-s)+\nu\,\Delta \gT(\vec z(s),t-s)\}\,ds
+\sqrt{2\,\nu}\vec \nabla \gT(\vec z(s),t-s)\cdot d\vec w(s)
\label{stodiff}
\end{eqnarray} 
This substitution is useful since the potential satisfies:
\begin{eqnarray}
\partial_{t}\gT +\frac{1}{2}\vec \nabla \gT \cdot\vec \nabla \gT &=&
\nu\Delta \gT \nonumber \\
\gT(\vec x,0)&=&\gT_{0}(\vec x) 
\label{scal}
\end{eqnarray} 
So we get to
\begin{equation}
\vec \nabla \gT(\vec x,t)\exp[-\frac{\gT(\vec x,t)}{2\,\nu}]=E^{\vec x}
\{\vec \nabla_{z(t)}\gT_{0}(\vec z(t))\exp[-\frac{\gT_{0}(\vec z(t))}
{2\,\nu}]\}
\end{equation} 
Finally, we can derive the explicit expression for the velocity potential 
by means of a simple integration by parts and by exploiting the homogeneity 
of (\ref{zeta}) :
\begin{equation}
\gT(\vec x,t)=-2\,\nu\,\ln E^{\vec x}\{\exp[-\frac{\gT_{0}(\vec z(t))}
{2\,\nu}]\}
\label{HC}
\end{equation} 
which means:
\begin{equation}
\gT(\vec x,t)=-2\,\nu\,\ln [\int_{-\infty}^{\infty}e^{-\frac{\Phi
(\vec x,\vec y,t)}{2\,\nu}}\,\frac{d^{d}y}{(4\,\pi\,\nu\,t)^{d/2}}]
\label{scalsol}
\end{equation}
where
\begin{equation}
\Phi(\vec x,\vec y,t)=\frac{(\vec x-\vec y)^{2}}{2\,t}+\gT_{0}(\vec y)
\label{phi}
\end{equation}
If we consider a class of initial conditions such that
\begin{equation}
\frac{\gT_{0}(\vec x)}{\|\vec x\|^{2}}\rightarrow 0 \quad \mbox{for} \quad
\|\vec x\| \uparrow \infty
\end{equation}
then (\ref{scalsol}) is always well defined and, in the limit 
$t \downarrow 0$, it is consistent with the initial condition $\gT(\vec x,0)=
\gT_{0}(\vec x)$. Therefore, we have recovered the already
known result of the Hopf-Cole theory (see e.g. \cite{Saichev2},
\cite{VeDuFrNo}) stating that the solution of the Burgers equation with 
irrotational initial condition is given in any dimension at arbitrary time 
$t$ by 
\begin{equation}
\vec v(\vec x,t) =\nabla \gT(\vec x,t)
\end{equation}

Let us now consider the system:
\begin{eqnarray}
\partial_{t}C +\vec v\cdot\vec \nabla C &=&\mu\,\Delta C +VC +g \nonumber \\
C(\vec x,0)& =& C_{0}(\vec x)
\label{dens}
\end{eqnarray}
where $\vec v$ is given by (\ref{buro}). The external drift $V$ and the volume 
force $g$ are functions of $\vec x$ and $t$; $V$, $g$ and the initial data 
$C_{0}$ are smooth functions growing, as $\|\vec x\|$ goes to infinity, 
more slowly than  $\|\vec x\|^{2}$. In the general case this equation 
can be formally integrated as a path integral in the form 
(see e.g. \cite{kara}):
\begin{eqnarray}
\lefteqn{C(x,t)=E^{\vec x}\{C_{0}(\vec z(t))\exp[-{\cal Z}_{t}+\int_{0}^{t}
V(\vec z(s),t-s)\,ds]\} +}\nonumber \\ 
&+& E^{\vec x}\{\int_{0}^{t}g(\vec z(s),t-s)\exp[-{\cal Z}_{s}+
\int_{0}^{s}V(\vec z(u),t-u)\,du]\,ds\} 
\label{solesem}
\end{eqnarray}
where $z(t)$ and ${\cal Z}_{t}$ are given respectively by (\ref{zeta}) and
(\ref{expsto}).

If $\mu=\nu$ and $\,V=g=0\,$\ the situation is more simple and the solution
can be easily expressed as an ordinary integral.
\begin{equation}
C(\vec x,t) = E^{\vec x}\{C_{0}(\vec z(t))e^{-{\cal Z}_{t}}\} 
\end{equation}
It is now sufficient to proceed, as before, to the elimination of the 
stochastic integral in (\ref{expsto}) by means of (\ref{stodiff}) and 
to use (\ref{scal}) to arrive to 
\begin{equation}
C(\vec x,t) = \exp[\frac{\gT(\vec x,t)}{2\,\nu}]\int_{-\infty}^{\infty}
C_{0}(\vec y)\exp[-\frac{\Phi(\vec x,\vec y,t)}{2\,\nu}]\,\frac{d^{d}y}
{(4\,\pi\,\nu\,t)^{d/2}}
\label{fs}
\end{equation}
or, in more explicit terms:
\begin{equation}
C(\vec x,t) = \frac{\int_{-\infty}^{\infty}C_{0}(\vec y)e^{-\frac{\Phi
(\vec x,\vec y,t)}{2\,\nu}}\,d^{d}y}{\int_{-\infty}^{\infty}e^{-\frac{\Phi
(\vec x,\vec y,t)}{2\,\nu}}\,d^{d}y}
\end{equation}
which is the first result of Saichev and Woyczynski.

If the diffusion coefficient in (\ref{dens}) is $\mu \ne \nu$ the velocity
and the concentration fields are, on the average, constant along different 
stochastic trajectories, so we can no more use (\ref{scal}) to reduce the path 
integral to a finite dimensional one. Still, we can formally write:
\begin{eqnarray}
C(\vec x,t)&=& e^{\frac{\gT(\vec x,t)}{2\,\mu}}E^{\vec x}
\{C_{0}(\vec z\,'(t))\exp[-\frac{\gT_{0}(\vec z\,'(t))}{2\,\mu}
-\frac{(\nu-\mu)}{2\,\mu}\int_{0}^{t}\Delta \gT(\vec z\,'(s),t-s)\,ds]\}
\nonumber \\ 
\vec z\,'(t)&\eql&{\cal N}(\vec x,2\,\mu\,t) 
\label{cfor}
\end{eqnarray}
where $\gT(\vec x,t)$ is given by (\ref{scalsol}). The expression (\ref{cfor})
holds true as far it is well defined (non divergent).

\section{A reaction-diffusion model}
\label{s:redm}

Another possible simple application of the Girsanov theorem is the solution 
of the following reaction diffusion model considered in \cite{Saichev}: 
\begin{equation}
\begin{array}{ccc}
\partial_{t}\vec v +\vec v \cdot\vec \nabla\vec v =\nu\Delta\vec v
+2\,\nu\,k \vec \nabla C  &\quad& \vec v(\vec x,0) =\vec \nabla \gT_{0}(\vec x)
 \label{rdmeq1}
\end{array}
\end{equation}
\begin{equation}
\begin{array}{ccc}
\partial_{t}C +\vec v\cdot\vec \nabla C =\nu\Delta C +k\,C^{2} &\quad&
C(\vec x,0) = C_{0}(\vec x) \label{rdmeq2}
\end{array} 
\end{equation} 
where $k$ is a constant. 

Let us consider the stochastic differential
\begin{eqnarray}
\lefteqn{d_{s}\{e^{k\int_{0}^{s}C(\vec x(u),t-u)\,du}\,C(\vec x(s),t-s)\} =} 
 \nonumber\\
&=&e^{k\int_{0}^{s}C(\vec x(u),t-u)\,du}\,\{k\,C^{2}(\vec x(s),t-s)\,ds + 
d_{s}C(\vec x(s),t-s)\}
\label{st2}
\end{eqnarray}
where
\begin{equation}
d\vec x(s) =- \vec \nabla \gT(\vec x(s),t-s)\,ds + \sqrt{2\,\nu}\,d \vec w(s) 
\end{equation}

If we take the expectation value of (\ref{st2}) and equation
(\ref{rdmeq1}) is satisfied, then it is easy to see that: 
\begin{equation}
C(\vec x,t) = E^{\vec x}\{C_{0}(\vec z(t))\exp[-{\cal Z}_{t}+ k
\int_{0}^{t}C(\vec z(s),t-s)\,ds]\} 
\end{equation}
where $z(t)$ and ${\cal Z}_{t}$ are given respectively by (\ref{zeta}) and
(\ref{expsto})

Again we can use (\ref{stodiff}) and we obtain an expression of the form 
(\ref{fs}) where, now,  $\gT(\vec x,t)$ is given by solving (\ref{rdmeq1}). 
This can be done by observing that, in terms of the potential, it has, 
according to the notation in \cite{Saichev}, the form:
\begin{equation}
\partial_{t}\gT +\frac{1}{2}\vec \nabla \gT \cdot\vec \nabla \gT = 
\nu\Delta \gT +2\,\nu\,k\, e^{\frac{\gT}{2\,\nu}}\,a_{0}  
\label{teta2}
\end{equation}
Here we have 
\begin{equation}
a_{0}(\vec x,t)=\int_{-\infty}^{\infty}C_{0}(\vec y)e^{-\frac{
\Phi(\vec x,\vec y,t)}{2\,\nu}}\,\frac{d^{d}y}{(4\,\pi\,\nu\,t)^{d/2}}
\label{a0}
\end{equation}
and $\Phi(\vec x,\vec y,t)$ is defined by (\ref{phi}).

The Ito calculus suggests that equation (\ref{teta2}) can be 
rewritten by means of the stochastic process (\ref{zeta}) as
\begin{equation}
E^{\vec x}\{d_{s}\exp[-\frac{\gT(\vec z(s),t-s)}{2\,\nu}]\}=k\,
E^{\vec x}\{a_{0}(\vec z(s),t-s)\}
\end{equation}
which implies
\begin{equation}
 \gT(\vec x,t)=-2\,\nu\,\ln E^{\vec x}\{e^{-\frac{\gT(\vec z(t),0)}
{2\,\nu}}-k\,t\,a_{0}(\vec z(t),0) + \int_{0}^{t}s\,d_{s}a_{0}
(\vec z(s),t-s)\,ds\,\}
\end{equation}
Since $a_{0}(\vec x,t)$ is solution of the heat equation we have
\begin{equation}
E^{\vec x}\{\int_{0}^{t}s\,d_{s}a_{0}(\vec z(s),t-s)\,ds\}=0
\end{equation}
Therefore, we can conclude that the velocity potential is now
\begin{equation}
\gT(\vec x,t)=-2\,\nu\,\ln[\,b_{0}(\vec x,t)-k\,t\,a_{0}(\vec x,t)]
\end{equation}
with
\begin{equation}
b_{0}(\vec x,t)=\int_{-\infty}^{\infty}e^{-\frac{\Phi(\vec x,\vec y,t)}
{2\,\nu}}\,\frac{d^{d}y}{(4\,\pi\,\nu\,t)^{d/2}}
\end{equation}
while the reactive tracer is
\begin{equation}
C(\vec x,t)=\frac{a_{0}(\vec x,t)}{b_{0}(\vec x,t)-k\,t\,a_{0}(\vec x,t)}
\end{equation}
which is the second result of Saichev and Woyczynski.

\section{Conclusion}
By means of Ito calculus is it possible to obtain the result 
reported by Saichev and Woyczynski in (\cite{Saichev}) in a natural and 
straight-forward way.

\section{Acknowledgements}
This work benefited from sharp observations made by Erik Aurell.
I am deeply grateful to him and to all the other friends in Sweden for
their warm hospitality and good time spent together.
This work was supported by a ESF/TAO grant for the year 1996.  

After the submission of the paper I received copy of the work by 
P. Garbaczewski et al. \cite{GarKon} where the Burgers equation is 
investigated in the more general Schr\"odinger interpolation framework.

\end{document}